\documentclass[twocolumn,aps,pra,showpacs,groupedaddress]{revtex4-1}
\usepackage[matrix,frame,arrow]{xy}
\usepackage{amsmath}

\usepackage{epsfig,amssymb,amsmath,mathrsfs,color} 
\usepackage[active]{srcltx} 
\usepackage[hypertex,linkcolor=red]{hyperref}
\usepackage[matrix,frame,arrow]{xypic}%\input{myqcircuit}
 \newtheorem{definition}{Definition}

 \def\>{\rangle}
\def\<{\langle}

\begin{document}
\title{Relativity of Causal Structure in Quantum Theory}
\author{Marco Zaopo}\email{marco.zaopo@unipv.it} 
\affiliation{Dipartimento di Fisica, Universit\`a
  di Pavia, via Bassi 6, 27100 Pavia, Italy}
\email{marco.zaopo@unipv.it}
\date{\today}
\begin{abstract}
Quantum theory is a mathematical formalism to compute probabilities
for outcomes happenning in physical experiments. These outcomes constitute
events happening in space-time. One of these events represents the fact that a system located in the region
of space where is situated a physical device has a certain
value of a physical observable at the time when the device fires
the outcome corresponding to that value of the observable. The causal
structure of these events is customarily assumed fixed in an
absolute way. In this paper
we show that this assumption cannot be substantiated on operational
grounds proving that two
observers looking at the same quantum experiment can calculate
the probabilities of the experiment assuming a different causal
structure for the space-time events
constituted by the outcomes. We will thus say that in quantum theory
we have relativity of causal structure.

\end{abstract}
\maketitle

In the standard, classical theory of probability, joint probability
distributions on the values of two random variables are defined
independently of the existence of a causal relationship between the
values of one of the variables and those of the other \cite{Defin}. This is the
case since a pair of random variables on which is defineable a joint
probability distribution can represent something out of the
domain of physics and thus not necessarily embedded in some given
space-time (we will refer to
space-time generically to a causal network of events, without
specifying any other property such as discreteness, continuity,
dimension etc.).
In quantum theory, on the contrary, two random variables always
represent two observables related to some physical system. The values
of these variables
are indeed associated to events embedded in some given space-time. In this case a given value of a
random variable is infact always percieved by a click of a detector
that has revealed a property of a physical system in some position of
space at some given instant of time. Hence the events on which are
defined joint probability distributions for two physical observables always
have a definite causal structure since any space-time ultimatley
constitutes a causal network of events. Quite recently, several authors have explored the
statistics of different quantum experiments in which are investigated the same
observables and phsyical systems and that differ in the causal
arrangements of the devices involved \cite{Leif1, Leif2, Brukvedr,
  EvPrWh, MarcRez}. In these papers it is
found that there are a lot of analogies from a formal point view in
describing two
quantum experiments differing only in the causal relations of the
devices involved. Exploiting this fact, in \cite{Leif1}, is
formulated quantum theory as a theory of Bayesian inference in which
the different causal relations between correlated regions are
treated in a unified way. In this paper we show that the mathematical
structure of quantum theory is such that two
observers looking at the same quantum experiment can calculate
the probabilities of the experiment assuming a different causal
structure for the events on which is defined the probability
distribution. This means that the causal structure of the events
happening in a quantum experiment may not be regarded as absolutely
fixed. Two observers can indeed obtain the information contained in a given
experiment assuming a different causal structure for the events
constituted by the outcomes involved. This result is likely to
have implications in the search of a theory of quantum gravity. It
suggests that to formulate a properly quantum theory of cosmological processes, we should look for a mathematical
formalism to calculate probabilities of these processes such that
the causal structure of the events on which is defined the probability
distribution of a process can be regarded as a mathematical symmetry. 
This paper is organized as follows. We first introduce a general framework
for quantum experiments whose information is contained in the
joint probabilities of the values of a pair of phsyical
observables. In this scenario, we formulate the property of relativity of causal
structure. We show that in any quantum experiment performable in the above framework we have relativity of causal
structure. We then discuss relativity of causal structure in relation
with the no-signalling principle. We finally link our result to those obtained in \cite{Leif1}.
\subsection{Operational framework}
In a generic quantum experiment one is interested in the joint
probabilities of the outcomes happening on two devices $\mathcal{A},
\mathcal{B}$, that are able to analyze a set of physical observables
generically pertaining
to two quantum systems $\mathscr{S}_1$ and $\mathscr{S}_2$ respectively
($\mathscr{S}_1, \mathscr{S}_2$ can of course be the same
system).
The possible values of a given observable
$A$ analyzed by device $\mathcal{A}$ constitute a set of outcomes of this
device $\{a_i\}_{i \in A}$ while the
possible values of another observable $B$, analyzed by device $\mathcal{B}$, constitute
another set of outcomes $\{b_j\}_{j
  \in B}$. The information contained in the experiment is expressed by
the probability distribution $\{p(a_i,b_j)\}_{i,j}$ for all $(a_i,b_j) \in
A\times B $ where it holds the normalization condition:
$
\sum_{i\in A, j \in B} p(a_i,b_j) = 1
$
This clearly holds for all the possible experiments
performable with devices $\mathcal{A}, \mathcal{B}$,
namely, for all the possible
observables that  $\mathcal{A}, \mathcal{B}$ can
analyze.
A simple example of this is an experiment involving two
Stern-Gerlach apparata SG$_1$, SG$_2$ analyzing the spin
of an electron. In this case observables $A$ and $B$ represent
two given orientations that the spin of an electron can have, for example
$Z_1$ and $Z_2$. The possible outcomes happening on SG$_1$ are $\{$spin
  up along $Z_1$ ($Z_1\uparrow$), spin down along $Z_1 (Z_1\downarrow) \}$ and those that can happen on
SG$_2$ are $\{$spin up along $Z_2 (Z_2\uparrow)$, spin down along $Z_2
(Z_2\downarrow)\}$. The information contained in the experiment is in
the joint probability distribution $\{p(Z_1\uparrow,Z_2\uparrow),
p(Z_1\uparrow,Z_2\downarrow), p(Z_1\downarrow, Z_2\uparrow),
p(Z_1\downarrow, Z_2\downarrow )\}$. 
The events on which are defined probability
distributions in quantum experiments always possess a definite causal
structure \cite{Wald}.
\begin{definition} \label{caustr}

Given a pair of events, $\chi_a, \chi_b$ it is defined a \emph{causal
structure} for these events if one of the following holds:
\begin{itemize}
\item $\chi_a$ causes $\chi_b$
\item $\chi_b$ causes $\chi_a$
\item $\chi_a$ does not cause $\chi_b$ and $\chi_b$ does not cause $\chi_a$
\end{itemize}
\end{definition}
This is the case since the random variables
on which the probability distributions are defined, refer to
observables pertaining to physical systems.
Consider an arbitrary pair of outcomes $(a_i,b_j) \in A\times B$ such that $p(a_i,b_j) \neq
0$. These outcomes constitute two events that can happen in
space-time. Outcome $a_i$ has associated an event $\chi_{a_i}$
representing that a system $\mathscr{S}_1$ has the value $a_i$
of observable $A$ in the region occupied by device
$\mathcal{A}$ with space coordinates $x_{a_i}$ at time $t_{a_i}$. In the same way outcome $b_j$ has
associated an event $\chi_{b_j}$ saying that $\mathscr{S}_2$ has
value $b_j$ of observable $B$ in a region occupied by $\mathcal{B}$ with space coordinates
$x_{b_j}$ at time $t_{b_j}$. To any quantum experiment is associated a specific dynamics of one or
more systems $\mathscr{S}$. Wether for the events $\chi_{a_i}, \chi_{b_j}$ associated to
the outcomes $(a_i,b_j)$ holds anyone of the alternatives in
definition \ref{caustr} clearly depends on the dynamics of the systems
involved in the experiment.
From an operational point of view, the assignment of a dynamics to the
systems in a quantum experiment 
consists of a specification of the inputs and outputs for the devices
involved in it. This point of view is first illustrated in \cite{hardy}. If we are interested in the joint
probabilities of outcomes happening on devices $\mathcal{A},
\mathcal{B}$, the possible input/output combinations assigned
to these devices are
responsible for the different alternatives in definition \ref{caustr} that can be associated to the pair of events
$\chi_{a_i}$, $\chi_{b_j}$ corresponding to the pair of outcomes
$(a_i,b_j)$. If the dynamics of
the experiment is such that system $\mathscr{S}_1$ is the
output of device $\mathcal{A}$ and $\mathscr{S}_2$ is the input of device
$\mathcal{B}$,
we have that the pair of events $\chi_{a_i}, \chi_{b_j}$
associated to $(a_i,b_j)$ are such that $\chi_{a_i}$ causes $\chi_{b_j}$. If, conversely, the dynamics is the time reversal of
the previous one, then system $\mathscr{S}_1$ is the input for
$\mathcal{A}$, system $\mathscr{S}_2$ is
the output for device $\mathcal{B}$ and the pair of events $\chi_{a_i}, \chi_{b_j}$
associated to $(a_i,b_j)$ are such that $\chi_{b_j}$ causes $\chi_{a_i}$. If the experiment is such that two causally independent systems
are inputs (or outputs) for devices $\mathcal{A}$ and $\mathcal{B}$, then the
causal structure of $\chi_{a_i}, \chi_{b_j}$
associated to outcomes $(a_i, b_j)$ is such that $\chi_{a_i},$ does
not cause $\chi_{b_j}$ and $\chi_{b_j}$ does not cause $\chi_{a_i}$,
namely, the two events are space-like. 
The assumption
that these input/output associations can be done in an absolute way can hardly be
motivated on operational grounds. There is infact no
experiment that can probe that a quantum system is ``escaped out
from a device'' in a given state and is ``entered into another
device'' causing an outcome happening on it. This is the case since
if it existed one such experiment, this should
also make the system interact with another probe system; the
interaction would perturb the dynamics of the original system and could in principle prevent it to
enter the aperture of a physical device or even to escape out from
it and would make the state and the measurement outcome change. A
similar reasoning on the impossibility to ``probe causal structrue''
in quantum theory can be found in \cite{hardy1}. In what follows we will infact show that two different
observers can compute the joint probabilities $p(a_i,b_j), \forall (a_i,b_j)
 \in A\times B$ in an experiment involving devices $\mathcal{A}$ and
 $\mathcal{B}$, assuming different input/output configurations
for these devices.
Since, from an operational point of view,
the specific causal structure of events $\chi_{a_i}, \chi_{b_j}$ associated to
the outcomes $(a_i,b_j)$ derives from the specification of the inputs and
outputs of the devices involved, we will say that in
quantum theory we have relativity of causal structure. 
\subsection{Relativity of Causal Structure}
Consider an arbitrary pair of outcomes $(a_i, b_j)$ having
non zero probability of jointly happening $p(a_i,b_j)$. 
An observer $O_{\alpha}$ assumes that a quantum system $\mathscr{S}_1$
is the output of device $\mathcal{A}$ on which $a_i$ happens, is
subject to an evolution $\mathscr{T}$ (eventually transforming $\mathscr{S}_1$ in
system $\mathscr{S}_2$) and then constitutes the input of a measurement
device $\mathcal{B}$ on which $b_j$ happens. This implies that the space-time events
$\chi_{a_i}, \chi_{b_j}$ associated to outcomes $a_i, b_j$ are assumed
such that $\chi_{a_i}$ causes $\chi_{b_j}$.
A second observer $O_{\beta}$ looking at the same quantum
experiment of $O_{\alpha}$ assumes that system $\mathcal{S}_2$ is the
output of device $\mathcal{B}$ where $b_j$ happens, is subject to an
evolution $\mathscr{T}'$ (eventually transforming $\mathscr{S}_2$ in
$\mathscr{S}_1$) and then constitutes the input of a measurement
device $\mathcal{A}$ on which it happens $a_i$. Since this constitutes
the time reversal of the dynamics assumed by observer $O_{\alpha}$,
space-time events $\chi_{a_i}, \chi_{b_j}$ are assumed by $O_{\beta}$
in such a way that $\chi_{b_j}$ causes $\chi_{a_i}$. 
A third observer,
$O_{\gamma}$, looking at the same experiment, is indeed assuming that systems $\mathscr{S}_1$ and
$\mathscr{S}_2$ are both causally independent inputs respectively of
two measurement devices $\mathcal{A}$
and $\mathcal{B}$ on which happen $a_i$ and $b_j$. The two systems
are both outputs of a preparation device for the composite system
$\mathscr{S}_1\mathscr{S}_2$ that prepares a state $\tau_{12}$. Observer
$O_{\gamma}$ thus assumes that
$\chi_{a_i}, \chi_{b_j}$ are two space-like events, namely,
$\chi_{a_i}$ does not cause $\chi_{b_j}$ and $\chi_{b_j}$ does not
cause $\chi_{a_i}$. 
\begin{definition}\label{rcs}
We have \emph{relativity of causal structure} if, given a
choice of mathematical objects performed by anyone of
$O_{\alpha}, O_{\beta}, O_{\gamma}$ to
calculate probability $p(a_i,b_j)$, there exist unique choices of
mathematical objects for the remaining two observers that permit them to calculate
$p(a_i,b_j)$. This must hold for all $(a_i,b_j) \in A\times B$ and for all $(A, B)$. 
\end{definition}
In what follows we will prove that in quantum theory we have
relativity of causal structure. 
Before doing this we will state a rule of transformation from mathematical objects
describing physical objects (i.e. evolutions, preparations and
measurement outcomes) used by an
observer $O$ to the corresponding mathematical objects used
by another observer $O'$.

$\bf{Transformation \;Rule}$ - 
\emph{Whenever a system $\mathscr{S}$ for which is defined a physical object
(i.e. preparation, evolution or measurement outcome) is seen as an input
(output) by observer $O$ and as an output (input) by observer
$O'$, the operator used to describe that object by $O$ is the transpose
on the Hilbert space of system $\mathscr{S}$ of the
operator used to describe the corresponding object seen by $O'$.}

Observer $O_{\alpha}$ 
assumes that $a_i$ is an element of a preparations ensemble
represented by a density matrix $\rho$ and a POVM $\{\bf{a_i}\}_{i \in A}$ such
that:
\begin{equation}\label{ro}
\rho = \sum_{i\in A} \text{Tr}[\bf{a_i}\rho] \frac{\sqrt{\rho} \;\bf{a_i} \sqrt{\rho}}{\text{Tr}[\bf{a_i}\rho]}
\end{equation}
It is easy to see that $\rho$ is a convex combination of density
operators for system $\mathscr{S}_1$ since $\sqrt{\rho} \bf{a_i} \sqrt{\rho} / \text{Tr}[\bf{a_i}\rho]$ is a positive
operator with unit trace defined on $\mathscr{S}_1$ for all $i$ while $\{\text{Tr}[\bf{a_i}\rho]\}_{i\in A}$
is a probability distribution. In what follows we assume $\rho$ not
pure, i.e. $i > 1$. We will discuss this assumption in section \ref{disc}. The transformation $\mathscr{T}$
transforming ensemble $\rho$ for observer $O_{\alpha}$ is represented
by a Completely Positive Trace Preserving (CPTP)
map $\mathscr{T}= \sum_{m} K^m\otimes
K^{m\dagger} $ where $K^m = \sum_{ab} K^m_{ab} |a\rangle_{2}
{}_{1}\langle b|$ is a Kraus operator \cite{kraus}. $b_j$ for $O_{\alpha}$ is a
measurement outcome represented by an element of a POVM
$\{\bf{b_j}\}_{j \in B}$ for system $\mathscr{S}_2$.
Using these informations and the rule stated above we can obtain
the mathematical objects used to describe the experiment by observers
$O_{\beta}$ and $O_{\gamma}$. In order to find the operators used to
describe the experiment by observer $O_{\beta}$, we note that he assumes input systems in correspondence of systems that
are outputs for
$O_{\alpha}$, while assumes output systems in correspondence of systems that
are inputs for
$O_{\alpha}$. Hence all the operators used by $O_{\beta}$ are the
transposed of those used by $O_{\alpha}$ since we have to transpose on
all spaces on which operators are defined. To find the operators used
by $O_{\gamma}$ we first explicitly write the evolution by means of
transformation $\mathscr{T}$ of ensemble $\rho$ seen by
$O_{\alpha}$. The density matrix obtained after the evolution by
$O_{\alpha}$ is:
\begin{equation}\label{tr}
\mathscr{T}({\rho}) = \sum_{m,ab,cd} K_{ab}^mK_{cd}^{m*}
|a\rangle_2 {}_1\langle b|\rho |c\rangle_1{}_2\langle d|
\end{equation}
Using the fact
that $\sum_{m} K^m\otimes
K^{m\dagger} $ can be written as:
\begin{equation} %\sum_{m,ab,cd} K_{ab}^mK_{cd}^{m*}
%|a\rangle_2 {}_1\langle b| |c\rangle_1{}_2\langle d| = 
\sum_{m,ab,cd} K_{ab}^mK_{cd}^{m*}
|c\rangle_1 {}_1\langle b| \otimes |a\rangle_2{}_2\langle
d|\end{equation} and the polar decomposition of $\rho$ we have:
\begin{equation}
\mathscr{T}({\rho}) = \text{Tr}_1[\sum_{m,ab,cd} K_{ab}^mK_{cd}^{m*}
\sqrt{\rho}|c\rangle_1 {}_1\langle b|\sqrt{\rho} \otimes
|a\rangle_2{}_2\langle d|]
\end{equation}
Note that, for the polar decomposition of $\rho$ to be uniquely
defined, one must assume $\rho$ to be full rank in the Hilbert space
corresponding to $\mathscr{S}_1$. This assumption is consistent with
the fact that we are considering that observer $O_{\alpha}$
describes the possible values of an observable $A$ as an ensemble of preparations representing the system
having all the different values of $A$. The density matrix obtained by
$O_{\alpha}$ after the evolution can thus be
written as $\mathscr{T}({\rho}) = \text{Tr}_1[\mathscr{T}_{\rho} ]$
where we define:
\begin{equation}\label{tr}
\mathscr{T}_{\rho} : = \sqrt{\rho} \otimes I_2 [\sum_{m} (K^m \otimes
K^{m\dagger})] \sqrt{\rho} \otimes I_2
\end{equation}
where $I_2$ is the identity matrix on system $\mathscr{S}_2$. From
(\ref{tr}) we see that
the evolution of ensemble $\rho$ can be
represented as an operator acting on Hilbert spaces of systems
$\mathscr{S}_1$ and $\mathscr{S}_2$. The evolution represented by 
$\mathscr{T}_{\rho}$ is seen as a bipartite state $\tau_{12}$ by $O_{\gamma}$ since he
assumes that the output of device $\mathcal{A}$ seen by $O_{\alpha}$,
$\mathscr{S}_1$, is indeed an input for $\mathcal{A}$. According to
the transformation rule stated above, $O_{\gamma}$ uses the
following mathematical object to represent the bipartite state
$\tau_{12}$:
\begin{equation}\label{t12}
\tau_{12} =  \mathscr{T}_{\rho}^{T_1} = \sqrt{\rho}^T \otimes I_2 [\sum_{m} (K^m \otimes
K^{m\dagger})^{T_1} ]\sqrt{\rho}^T \otimes I_2  
\end{equation}
Where ${}^{T_1}$ denotes partial transpose on space 1 corresponding to
$\mathscr{S}_1$. To see that (\ref{t12}) is a normalized bipartite
state we define the normalized bipartite state on two copies of $\mathscr{S}_1$, $|\Phi\rangle_{11'}$:
\begin{equation}\label{fi}
|\Phi\rangle_{11'} = \sqrt{\rho}^T \otimes I_{1'} 
\sum_j | j \rangle_{1} \otimes |j \rangle_{1'}
\end{equation}
where $\{|j\rangle\}_{j=1}^{d_1}$ is an orthonormal basis for Hilbert space of system
$\mathscr{S}_1$. 
Exploiting (\ref{fi}) we can write:
\begin{equation}\label{isot}
\mathscr{I} \otimes \mathscr{T} (|\Phi\rangle\langle\Phi|) = \tau_{12}
\end{equation}
where $\mathscr{I}$ is the identity map on system
$\mathscr{S}_1$ and $\mathscr{T}$ represent the evolution defined
above. From (\ref{isot}) we can see that $\tau_{12}$ is a
normalized bipartite state since $\mathscr{T}$ is a TPCP map acting on
system $\mathscr{S}_1$ and $|\Phi\rangle\langle\Phi|$ is a normalized
bipartite state.
The element of ensemble of preparations corresponding
to $a_i$ for $O_{\alpha}$ is seen by $O_{\gamma}$ as a measurement
outcome. In consequence of this it is represented as
$\bf{a_i^T}$ by $O_{\gamma}$ since he assumes system $\mathscr{S}_1$
as an input contrary to $O_{\alpha}$.  
The probability $p_{\alpha}(a_i,b_j)$ calculated by observer $O_{\alpha}$ is:
\begin{equation}\label{oalfa}
p_{\alpha}(a_i,b_j) = \text{Tr}_2[\bf{b_j} \text{Tr}_1[\mathscr{T}_{\rho} \bf{a}_i]]
\end{equation} 
The probability calculated by $O_{\beta}$ is:
\begin{equation}\label{obeta}
p_{\beta}(a_i,b_j) = \text{Tr}_1[\bf{a_i}^T \text{Tr}_2[\mathscr{T}_{\rho}^T \bf{b}_j^T]] 
\end{equation}
The probability calculated by $O_{\gamma}$ is:
\begin{equation}\label{ogamma}
p_{\gamma}(a_i,b_j) = \text{Tr}_{12} [\bf{a_i}^T \otimes \bf{b_j} \mathscr{T}_{\rho}^{T_1}]
\end{equation}
It can be easily verified that $p_{\alpha}(a_i,b_j) = p_{\beta}(a_i,b_j) =
p_{\gamma}(a_i,b_j)$. Since given an operator corresponding to
a preparation, transformation or measurement outcome seen by a given
observer, its transpose and its partial transpose
on any of its subspaces are uniquely defined and we assumed $(a_i,b_j)$
arbitrary, we have
relativity of causal structure by definition \ref{rcs}. 
\subsection {Discussion and related work}\label{disc}
First we discuss the assumption done after (\ref{ro}) that $\rho$ is not
a pure state. This is necessary for $\tau_{12}$ in (\ref{t12}) to be
normalized. If we have $\rho = |a\rangle\langle
a|$ then observer $O_{\alpha}$ is interested only in joint
probabilities of the type $p(a, b_j)$ with $b_j \in \{b_j\}_{j \in B}$
and $a$ fixed value of
observable $A$. This is equivalent to state that the uncertainty in
observable $A$ is 0 and $p(a, b_j) = p(b_j|a)$. Observers
$O_{\gamma}$ and $O_{\beta}$ clearly cannot assume that the
uncertainty in $A$ is 0, since from their point of view this
represents a measurement whose outcomes are random.
Nevertheless they can compute the
above probability as 
\begin{equation}\label{relat}
p(b_j|a) = p(a,b_j)/\sum_{b_j}p(a,b_j). 
\end{equation}
This is the fraction of times $b_j$ happens given that $a_i = a$ has happened and
is the probability obtained by $O_{\alpha}$ assuming $\rho = |a\rangle\langle
a|$. The probability $p(a,b_j)$ written in (\ref{relat}) for
$O_{\beta}$ and $O_{\gamma}$ can be calculated for an arbitrary
probability distribution on the values $\{a_i\}_{i\in A}$ of observable
$\mathcal{A}$. 
We should now discuss the relationship between relativity of causal structure
and the ``no-signalling principle''.
\begin{definition}\label{nosig} \emph{No-signalling principle}

If two devices $\mathcal{A}$ and
$\mathcal{B}$ are space-like separated originating
outcomes corresponding to space-like events, then 
\begin{equation}\label{ns}
p(b_j) = \sum_{a_i} p(a_i,b_j) = \sum_{a'_i} p(a_i',b_j) \;\;\;\forall\;\;
\{a_i\}_{i\in A} , \{a_i'\}_{ \in A'}
\end{equation}  
for all $b_j \in \{b_j\}_{j \in B}$ where $B$ is a measurement on
device $\mathcal{B}$ and $A$ and $A'$ are any two different measurements performed on
device $\mathcal{A}$. 
\end{definition}
Since this principle holds in quantum theory, the quantum correlations between
space-like devices cannot be used by an agent
operating on $\mathcal{B}$ to become aware of the actions of an agent
operating on device $\mathcal{A}$. Note that
(\ref{ns}) is only a necessary condition that joint probabilities of
outcomes happening on two space-like separated devices must
satisfy. Hence
there are no contradictions for an observer $O_{\alpha}$ assuming
that for every pair of
outcomes $(a_i,b_j) \in A\times B$ the associated space-time events
$\chi_{a_i}, \chi_{b_j}$ are such that, say, $\chi_{a_i}$ causes
$\chi_{b_j}$. On the other hand, if $\sum_{a_i} p(a_i,b_j) \neq \sum_{a'_i} p(a_i',b_j)$ for some $b_j$ and a pair
$\{a_i\}_{i\in A} , \{a_i'\}_{ \in A'}$ then an observer $O_{\alpha}$
establishes that an agent
operating on $\mathcal{A}$ must have changed the ensemble of preparations
from $\rho = \sum_{i\in A} \text{Tr}[\bf{a_i}\rho] \sqrt{\rho}
\;\bf{a_i} \sqrt{\rho}/\text{Tr}[\bf{a_i}\rho]$ to
$\rho' = \sum_{i'\in A} \text{Tr}[\bf{a_i'}\rho] \sqrt{\rho'} \;\bf{a_i'} \sqrt{\rho'}/\text{Tr}[\bf{a_i'}\rho']$. An observer
$O_{\gamma}$ looking at the same experiments establishes that
correlations on devices $\mathcal{A}$ and $\mathcal{B}$ are in one
case due to a
bipartite state $\tau = \mathscr{T}_{\rho}^{T_1}$ (with $\mathscr{T}$
evolution of ensemble $\rho$ seen by $O_{\alpha}$ and
$\mathscr{T}_{\rho}$ defined in (\ref{tr})) and in the other
case to a bipartite state $\tau' = \mathscr{T}_{\rho'}^{T_1}$; in
both cases ${}^{T_1}$ means transposition on Hilbert space on
which $\rho, \rho'$ are defined.

In (\ref{t12}) it is introduced an isomorphism between bipartite states
and evolutions of preparations ensembles via partial transposition of
the corresponding operators. This isomorphism is also introduced in
\cite{Leif1} where it is
invented the formalism of \emph{quantum conditional states.} Quantum
conditional states are used to formulate a theory of Bayesian
inference for random variables representing physical observables
pertaining to two regions that have a definite causal
relationship. The peculiarity of this theory is a tool called
\emph{star product}. Star product 
permits to perform statistical inference for two correlated regions
A and B
in strict analogy with the ordinary theory of probability in which
there is no dependence on the causal relationship between the regions.
Quantum conditional states are divided into causal conditional states and
acausal conditional states depending on wether the two regions are
causally related regions (an outcome in one region causes that in the
other region or viceversa) or not. A CPTP map, $\mathscr{T}_{AB}$ from
region A to region B, is
related to an \emph{acausal} conditional state $\rho_{A|B}^s$, by
means of the
Choi isomorphism \cite{choi}:
\begin{equation}
\mathscr{T}_{AB} \leftrightarrow \mathscr{I}_{A'} \otimes \mathscr{T}_{A''B} (|\Phi^+\rangle\langle\Phi^{+}|) = \rho_{\text{A}|\text{B}}^s
\end{equation} 
where $|\Phi^+\rangle = \frac{1}{\sqrt{d_A}}\sum_i | i \rangle_{A'}
\otimes |i \rangle_{A''}$ and $\{|i\rangle\}_{i=1}^{d_A}$ is a basis for
Hilbert space pertaining to the system in region A and $A'$, $A''$ two
copies of the system in region A. The rule of belief propagation
is used to find the joint state $\rho_{AB}^s$ for two systems
in space-like
separated regions, A and B, starting from the prior
pertaining to one of the two regions, $\rho_A$; this is expressed
via the star product:
\begin{equation}
\rho_{AB}^s = \rho_A\star \rho_{A|B}^s = d_A \sqrt{\rho_A} \otimes I_B \; \rho_{A|B}^s
\;\sqrt{\rho_A} \otimes I_B
\end{equation}
The star product used here involves also a
normalization factor $d_A$ that cancels with the factor $1/d_A$
arising from the definition of conditional state involving $|\Phi^+\rangle$.
Note that $\rho_{AB}^s$ is equal to $\tau_{12}$ defined in (\ref{t12})
with $\mathscr{S}_1$ and $\mathscr{S}_2$ pertaining to devices
$\mathcal{A}$ and $\mathcal{B}$ respectively in
regions A and B.
The map $\mathscr{T}_{AB}$ is related to a \emph{causal} conditional
state $\rho_{A|B}^t$ by means of the Jamiolkowsky isomrphism \cite{jamio}:
\begin{equation}
\mathscr{T}_{AB} \leftrightarrow [\mathscr{I}_{A'} \otimes \mathscr{T}_{A''B}
(|\Phi^+\rangle\langle\Phi^{+}|)]^{T_{A'}} = \rho_{A|B}^t 
\end{equation}
where ${}^{T_{A'}}$ denotes partial transposition on Hilbert space of
system $A'$
pertaining to region A. The rule of belief propagation
is used to find the joint state $\rho_{AB}^t$ for two systems in two causally related
regions A and B (or equivalently for one system at two
different times) starting from the prior
pertaining to region A, $\rho_A$. This is expressed with the star
product as above:
\begin{equation}
\rho_{AB}^t = \rho_A^T \star \rho_{A|B}^t =  d_A \sqrt{\rho_A^T} \otimes I_B \; \rho_{A|B}^t
\;\sqrt{\rho_A^T} \otimes I_B
\end{equation}where ${}^{T}$ denotes transposition.
Note that here $\rho_{AB}^t$ is equal to $\mathscr{T}_{\rho}$ defined
in (\ref{tr}) with $\mathscr{S}_1$ and $\mathscr{S}_2$ pertaining to devices
$\mathcal{A}$ and $\mathcal{B}$ respectively in
regions A and B.

\subsection{Conclusion}
In conclusion we have showed that the assumption of an absolute causal
structure for the space-time events associated to the outcomes in a quantum experiment cannot be motivated on operational
grounds. We infact showed that two observers looking at the same
quantum experiment can compute the relevant probabilities
assuming a different causal structure for the events on which is
defined the probability distribution. This can have implications in quantum gravity. In light of this
result, a possible way to concieve a theory of quantum gravity is to
look for a formalism to compute probabilities of cosmological
processes such that the causal structure of the events on which is
defined the probability distribution of a process can be regarded as a mathematical
symmetry.

\end{document}